\documentclass[a4paper,twoside,10pt]{letter}
\usepackage{amssymb}
\usepackage{color}
\usepackage{graphicx,saj,multicol,subeqnarray}

\def\papertype{Original Scientific Paper}

\begin{document}
\baselineskip=3.1truemm
\columnsep=.5truecm
\newenvironment{lefteqnarray}{\arraycolsep=0pt\begin{eqnarray}}
{\end{eqnarray}\protect\aftergroup\ignorespaces}
\newenvironment{lefteqnarray*}{\arraycolsep=0pt\begin{eqnarray*}}
{\end{eqnarray*}\protect\aftergroup\ignorespaces}
\newenvironment{leftsubeqnarray}{\arraycolsep=0pt\begin{subeqnarray}}
{\end{subeqnarray}\protect\aftergroup\ignorespaces}
%


\markboth{\eightrm EMISSION ACTIVITY OF THE BE STAR 60 CYGNI} {\eightrm K. \v{S}EJNOV\'{A}, V. VOTRUBA}

{\ }

\publ

\type

{\ }


\title{Emission activity of the Be star 60 Cygni}


\authors{K. \v{S}ejnov\'{a}$^{1,2}$, V. Votruba$^{1,2}$ }

\vskip3mm


\address{$^1$Department of Theoretical Physics and Astrophysics, Kotl\'{a}\v{r}sk\'{a} 2, 611 37 Brno, Czech Republic}	
\address{$^2$Astronomical Institute of the Academy of Sciences, 251 65 Ond\v{r}ejov, Czech Republic}
\Email{k.sejnova}{gmail.com, votruba@physics.muni.cz}


\dates{}{}


 \summary{In this paper we present results of the spectroscopic analysis of H$\alpha$ line profile of the Be star 60 Cygni. We present time evolution of the equivalent width of the H$\alpha$ line profiles during years 1992 - 2016 and $V/R$ variation during years 1995 - 2016. We analyzed data from Ond\v{r}ejov Observatory and from BeSS Database. The circumstellar disk of the star was present twice during years 1992 - 2016 and the second cycle shows stronger emission activity. We found out that the formation of the disk takes longer time than the disk extinction (extinction is much steeper than the formation) and that there is no evident period of changes in $V/R$ variation.}

\keywords{Stars: emission-line, Be -- Stars: individual: 60 Cygni.}

\begin{multicols}{2}
{


\section{1. INTRODUCTION}

In general Be stars are rapidly rotating B type stars that produce a disk in its equatorial plane and whose spectrum have or had at some time, one or more Balmer lines in emission. The definition of the Be stars indicates that the variability is very important in understanding the Be stars. Be stars show several types of variabilities, from long-term variations which take several years  to short-term variations which can take  only a few days or even minutes. Long-term variations are connected with an appearance and disappearance of the disk that means presence of the emission lines - phase changes of stellar types: B $\leftrightharpoons$ Be, Be $\leftrightharpoons$ Be-shell, and B $\leftrightharpoons$ Be-shell in time scales from a few years up to several decades. These variations are common for all the Be stars (Kogure \& Leung 2007).
\par For a long time, the star 60 Cyg (V1931 Cyg, HD 200\,310, HR 8\,053, MWC 360; B1 Ve, $V =5{.}37\,\mathrm{m}$ (var.), $v\,\mathrm{sin}\,i = 320\,\mathrm{km}\,\mathrm{s}^{-1}$), according to Hoffleit \& Jaschek (1982), has been known as an emission-line star. This star was studied by several authors (e.g. Koubsk\'{y} et al. (2000), Wisniewski et al. (2010), Draper et al. (2011) or Draper et al. (2014)).
\par Koubsk\'{y} et al. (2000) concluded that long-term variations, both spectroscopic and photometric, are indicative of a gradual formation and dispersal of the Be envelope around 60 Cyg.  Medium and rapid time-scales changes were found as well. Periodical radial velocity variations of spectral lines H$\alpha$ and He I $6678\,\mathrm{\AA{}}$ suggest that the star might be a spectroscopic binary having a period of $146{.}6\pm0{.}6\,\mathrm{days}$, while $1{.}064\,\mathrm{day}$ line profile variations and $0{.}2997\,\mathrm{day}$ photometric variations may be caused by non-radial pulsations. 
\par Wisniewski et al. (2010) presented one of the most comprehensive spectropolarimetric view of transition from Be phase to the normal B-star phase to date. They presented 35 spectropolarimetric and 65 $\mathrm{H}\alpha$ spectroscopic observations of Be star 60 Cyg spanning 14 years. They found that the timescale of the disk-loss events in 60 Cyg corresponds to almost 6 complete orbits of star's binary companion. This suggest that star's binary companion does not influence the primary star (or its disk). From Wisniewski et al. (2010) we know that the position angle of intrinsic polarization arising from 60 Cyg's disk is $\theta_\ast=107{.}7^{\circ}\pm0{.}4^{\circ}$ indicating that the disk situated in the equatorial plane is oriented on the sky at a position angle of $\theta_{\mathrm{disk}}=17{.}7^{\circ}$ (measured North to East).  The star was also studied by Draper et al. (2011) who analyzed the intrinsic polarization in the process of losing its circumstellar disk via Be to normal B star transition.

\section{2. OBSERVATIONS}

We used 38 spectroscopic observations of 60 Cygni. Spectra were observed by Perek telescope at Ond\v{r}ejov Observatory in the period from 2003 to 2011. This work also made use of the BeSS Database, operated at LESIA, Observatoire de Meudon, France: http://basebe.obspm.fr. We used all the available data for the star 60 Cyg. These data were observed by several astronomers in years between 1995 and 2016. We noted observers for each spectrum to be easily recognized if reader would like to see the spectra from the BeSS Database.

\section{3. RESULTS}

\subsection{3.1. Equivalent width}

The equivalent width of a spectral line ($EW$) is a measure of the area of the line on a plot of intensity versus wavelength. It is found by forming a rectangle with a height equal to that of continuum emission, and finding the width such that the area of the rectangle is equal to the area in the spectral line (Carroll \& Ostlie (2007)). Positive value of $EW$ is commonly used for absorption line  and negative value for  emission line, we use the same notification.
\par We used SPEFO code (developed by Ji\v{r}\'{\i} Horn, see \v{S}koda 1996) to analyze spectra from Ond\v{r}ejov Observatory and from BeSS Database and to measure the equivalent width of H$\alpha$ line profiles ($EW$(H$\alpha$)). Resultant values for $EW$(H$\alpha$) of spectra observed at Ond\v{r}ejov Observatory can be seen in Table~1. In Fig.~1 we can see the time evolution of the EW(H$\alpha$) for these data (empty circles). In Table~2 values for $EW$(H$\alpha$) for spectra adopted from BeSS Database are presented, values are presented by black circles in Fig.~1. 
\par In Fig.~1 we also plotted values of $EW$(H$\alpha$) from work by Wisniewski et al. (2010) (presented by gray triangles).  They used observations from years between 1992 and 2006 and all the data from the article can be found online.
\centerline{\includegraphics[width=1\columnwidth, keepaspectratio]{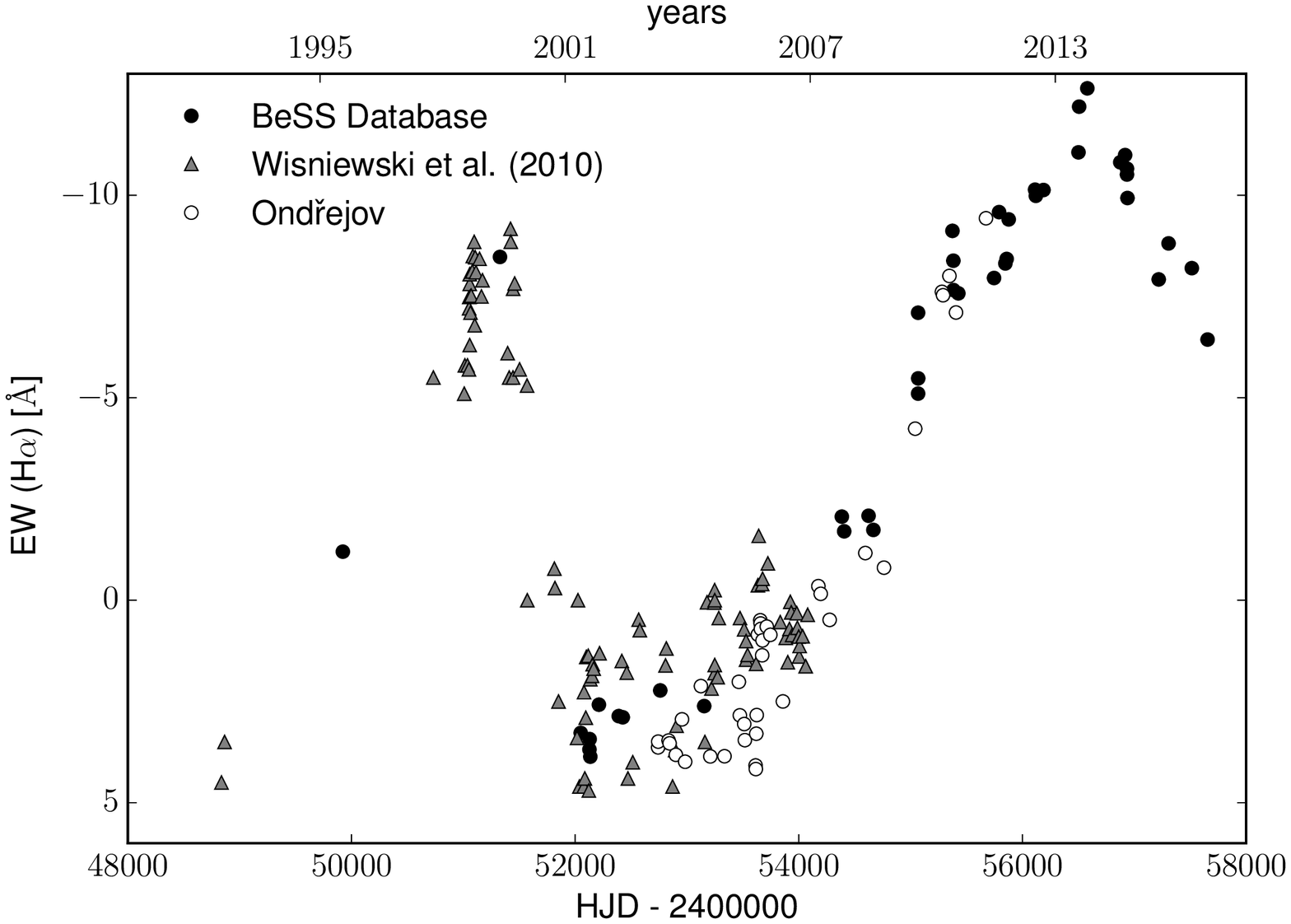}}
\figurecaption{1.}{Evolution of a equivalent width of H$\alpha$ line in \AA{}. Empty circles mark values for spectra observed at Ond\v{r}ejov Observatory, gray triangles mark values from Wisniewski et al.~(2010) and black circles mark the spectra adopted from BeSS Database but EW value were determined in this work. }}
\par As it is assumed the higher the value of the $EW$(H$\alpha$) the bigger the size of the circumstellar disk and for pure absorption there is no disk present. If we define that one cycle of the ``life'' of the disk is from the lowest state of pure absorption through the maximum strength of the H$\alpha$ emission and back to the lowest state we can say from Fig.~1 that between years 1992 and 2016 the disk was presented twice. The second cycle is not finished since there is no further observation of the star (the last observation is from $23^{\mathrm{rd}}$~September~2014) and thus we cannot be sure if the value of $EW$(H$\alpha$) would continue to increase or decrease. It is obvious that in the second cycle the 60 Cyg shows stronger emission activity.  
\par Furthermore, Koubsk\'{y} et al. (2000) noted that observations from years 1953 -- 1999 seem to indicate that the maximum strength of the H$\alpha$ emission never exceeds a peak intensity of about 2.0 of the continuum level but our new observations and observations from BeSS Database show several profiles with intensity over 2.0. So it seems that 60 Cygni is showing the strongest emission activity in the last 60 years.
\par Wisniewski et al. (2010) presented that their data indicate that it took 870 days for 60 Cyg's disk to transform from its strongest H$\alpha$ emission state to its lowest state of pure absorption, the latter of which they interpret as the time when the disk had completely dissipated. If we would do the same study for the formation of the disk we notice that it takes longer time. That means that times for formation of a disk and for dissipation of a disk are not the same.
\par Nevertheless from the Fig.~1 it is evident that a  periodicity in the evolution of value of $EW$(H$\alpha$) can be found. This work made use of python module pdm.py  to determine the period. This module is patterned on the method of period determination using phase dispersion minimization (Stellingwerf (1978)). Phase dispersion minimization (PDM) is a data analysis technique that searches for periodic components of a time series data set. It is useful for data sets with gaps, non-sinusoidal variations, poor time coverage or other problems that would make Fourier techniques unusable. 
This method provides likelihood estimate in the form of theta where lower theta represents higher certainty of prediction. 
See Fig.~2 for the result of the analysis, the procedure found period $6009\pm52\,\mathrm{days}$ (period for the minimal theta ($\theta=0.15$) is 6050{.}1\,days). pdmEquiBin carry out the PDM analysis using equidistant bins and pdmEquiBinCover carry out the PDM analysis using multiple sequences of equidistant bins.
\par We also used Period04 (Lenz \& Breger (2005)) for period analysis and we found  value for the period close to the one found by pdm.py: $P=6172{.}8\pm86{.}4\,\mathrm{days}$, see Fig.~3. But in Fig.~4 we can see that the Least-Sqaures Fit does not fit data well. For better understanding of this dependency it would be convenient to watch the star closely in the future and gain more data from the past. 
\centerline{\includegraphics[width=1\columnwidth, keepaspectratio]{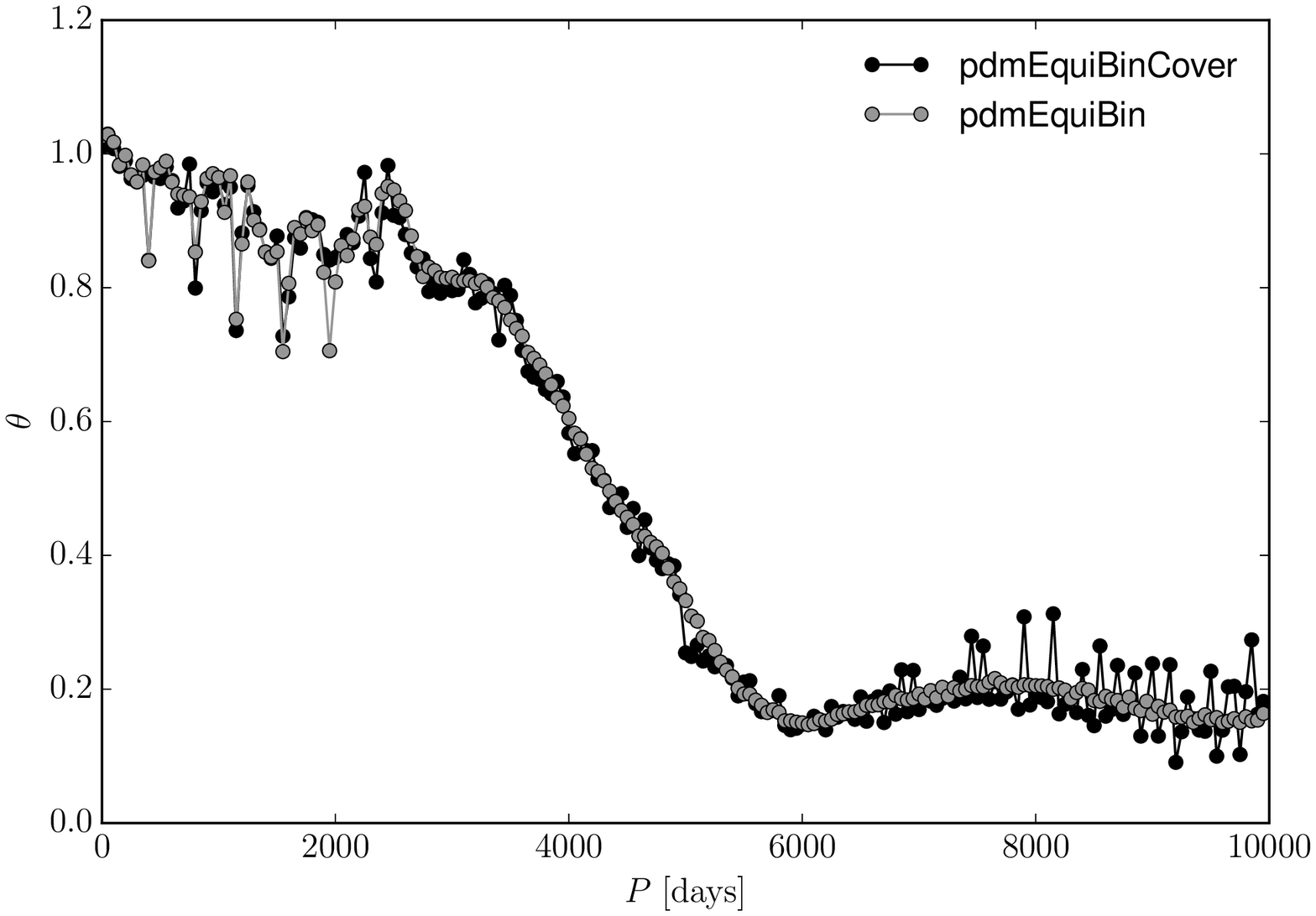}}
\figurecaption{2.}{Results of PDM analysis: $P=6009\pm52\,\mathrm{days}$ }
\centerline{\includegraphics[width=1\columnwidth, keepaspectratio]{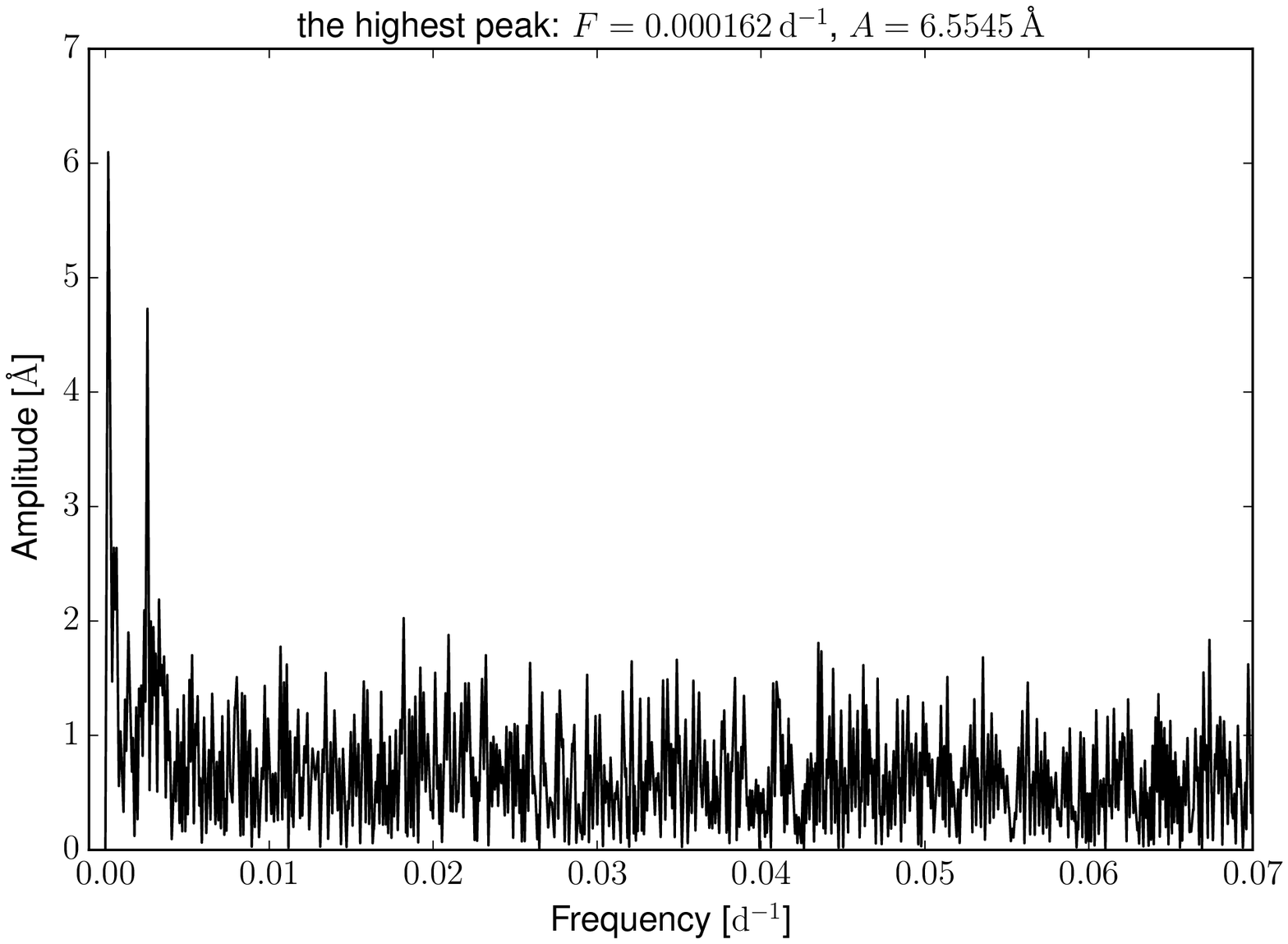}}
\figurecaption{3.}{Results of period analysis using Period04, where $F$ is the frequency, $A$ is the amplitude. Value of found period is $P=6172{.}8\pm86{.}4\,\mathrm{days}$.}
\centerline{\includegraphics[width=1\columnwidth, keepaspectratio]{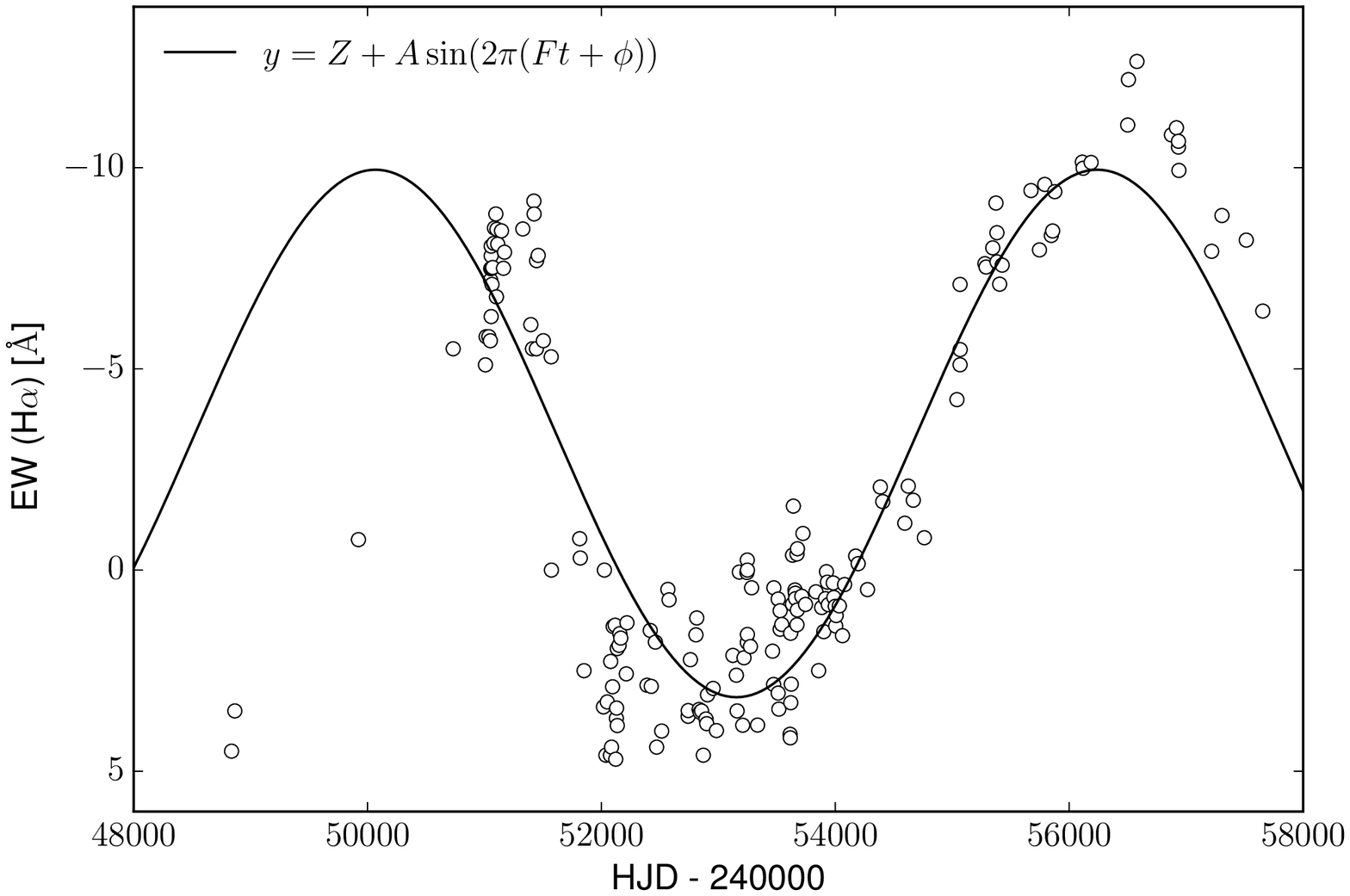}}
\figurecaption{4.}{The Least-Squares Fit for data gained by Period04. $Z=-3.39$ as the zero point, $A=6.56\,\mathrm{\AA{}}$ as the amplitude, $F=0.000162\,\mathrm{d^{-1}}$ as the frequency, $\phi=0.639$ as the phase and $t$ as the time. }
\section{3.2 \boldmath{$V/R$} VARIATION}

When the Be stars have double-peaked emission lines we usually observe $V/R$ (violet/red) variations. $V/R$ variation is expressed as the ratio of respective emission-peak heights above the underlaying photospheric absorption profile. Long-term cyclic changes in the ratio $V/R$ are observed in many stars, taking from a few years up to decades to complete the cycle and we can observe this phenomena in 60 Cygni spectra as well. We created a~simple script to determine values for $V$ and $R$ peaks of H$\alpha$ line profiles. Values for the $V/R$ and $(V+R)/2$ variations are presented in Table~1 and Table~2, for data from Ond\v{r}ejov Observatory and BeSS Database, respectively.
\par In Fig.~5 we plotted time evolution of $V/R$ ratio of H$\alpha$ emission line profiles.   We also used Period04 for the period analysis for $V/R$ variation, see Fig.~6. From the graph we can say that there is no evident period in the time evolution of data for this variation. 
The values of $V/R$ variation fluctuate in range $0{.}96-1{.}06$. It can be due to the fact that there is no one-armed oscillation of the disk present or it is very small and does not influence the $V/R$ variations (one-armed oscillation of the disk is one of the theory that is trying to explain the $V/R$ variation).
\par Time evolution of variation of $(V+R)/2$ for  H$\alpha$ emission line profiles can be seen in Fig.~7. In both graphs we plotted $V/R$ and $(V+R)/2$ values for H$\alpha$ line profiles observed at Ond\v{r}ejov Observatory (empty circles) and those adopted from BeSS Database (black circles). 

\centerline{\includegraphics[width=1\columnwidth, keepaspectratio]{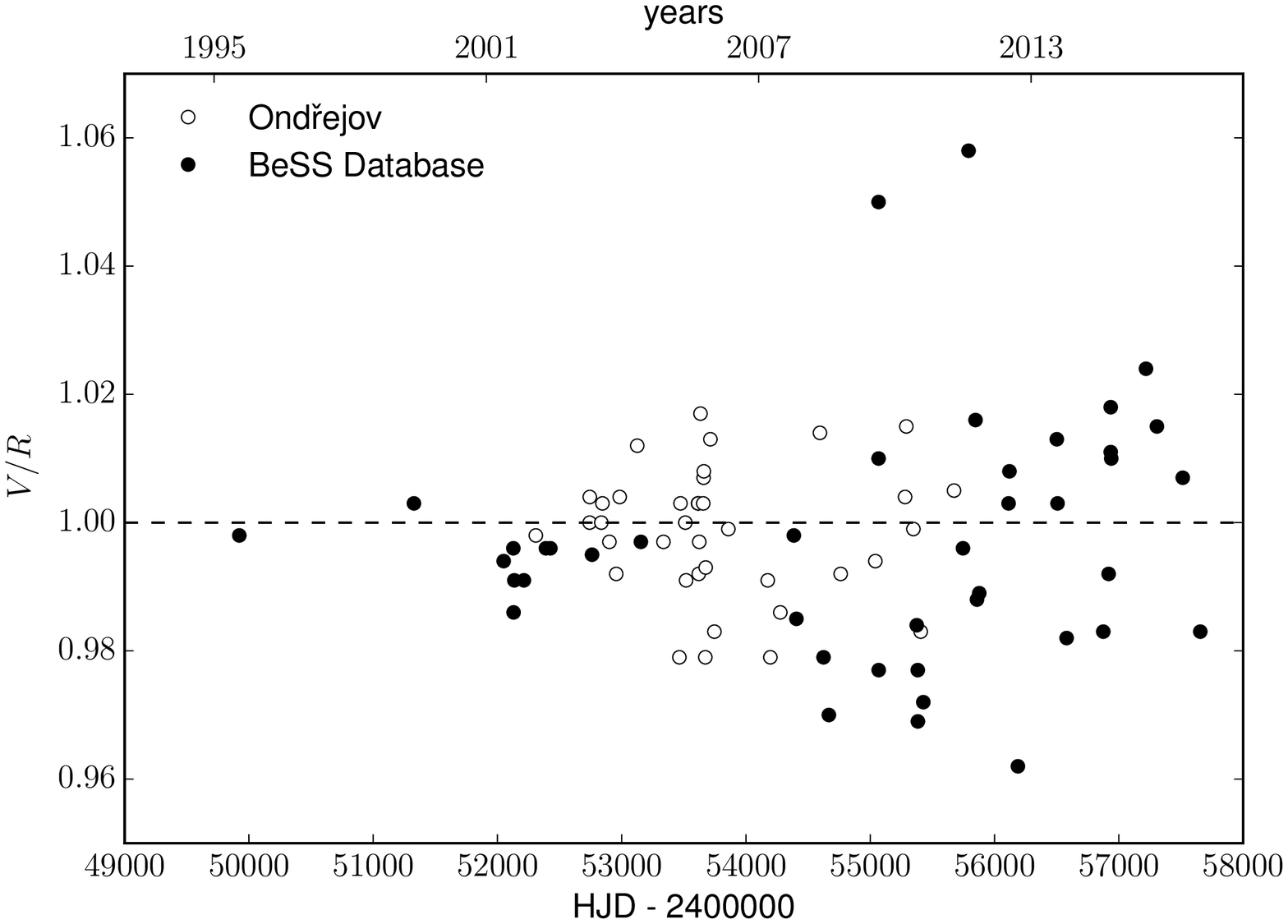}}
\figurecaption{5.}{$V/R$ ratio of H$\alpha$  line profile, dashed line marks normalized intensity with value 1.}
\centerline{\includegraphics[width=1\columnwidth, keepaspectratio]{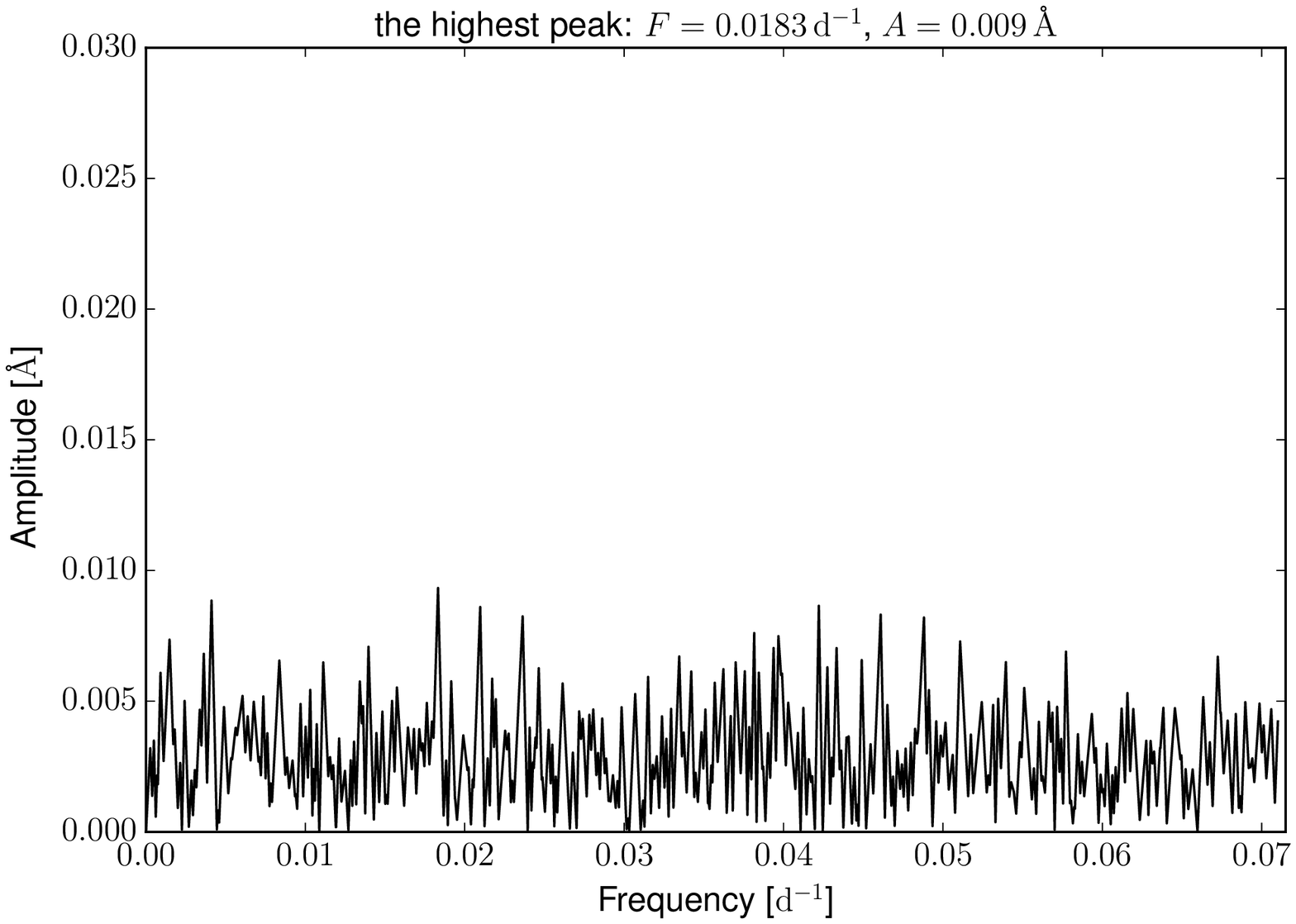}}
\figurecaption{6.}{Results of period analysis for $V/R$ variation using Period04.}
\par From Fig.~2 in Koubsk\'{y} et al. (2000) we can see $V/R$ and $(V+R)/2$ ratios between years 1976 to 1999 for Be star 60 Cygni. It is clear to see that the $(V+R)/2$ has its peak around 1.75 (at HJD $\approx$ 51000) and then starts decreasing. This peak has smaller value than the one we determined from our graph in this work which is around 2{.}15 (at HJD $\approx$ 57000). We can see the same behavior in development of $EW$(H$\alpha$) time evolution.
\centerline{\includegraphics[width=1\columnwidth, keepaspectratio]{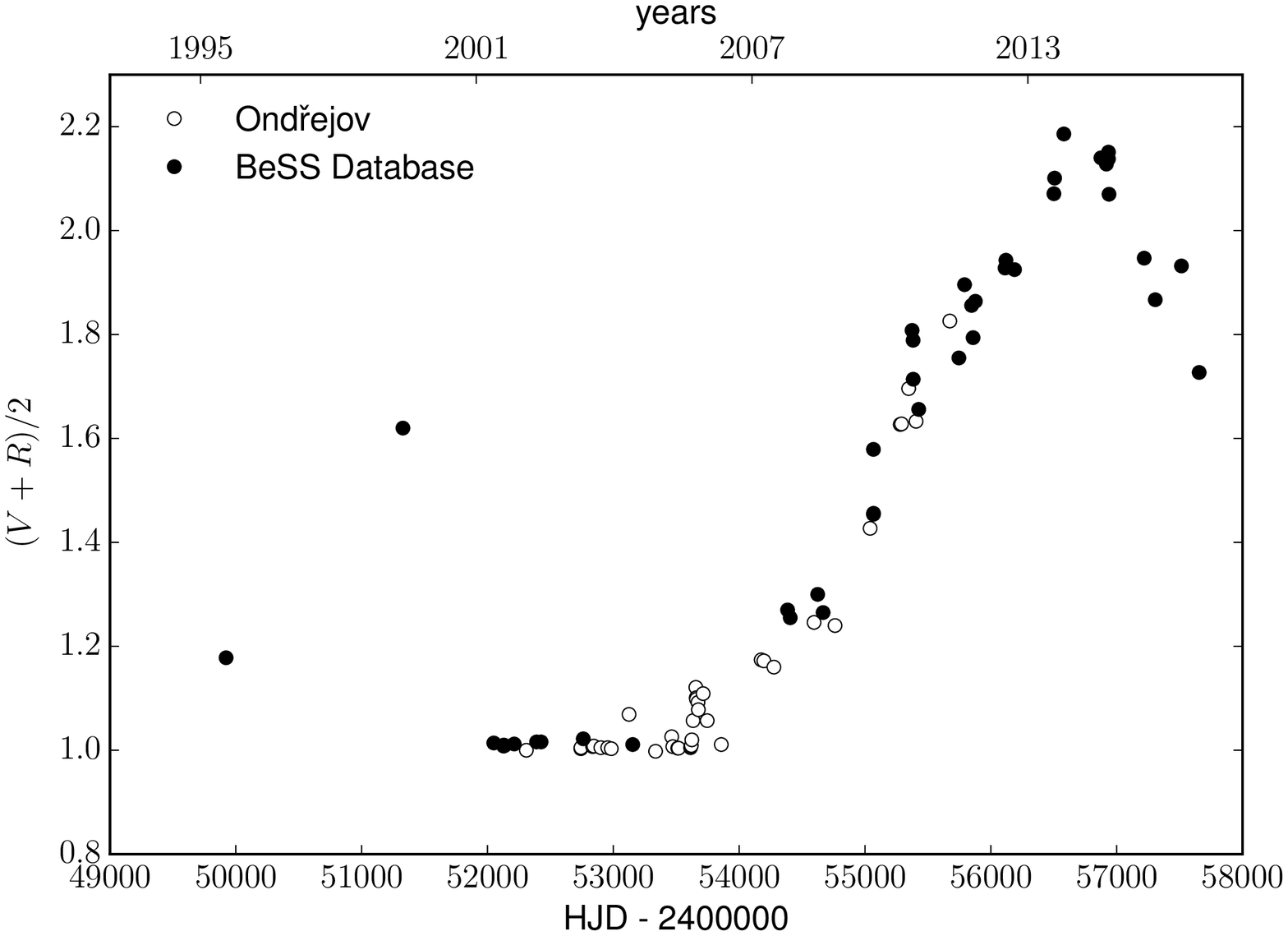}}
\figurecaption{7.}{$(V+R)/2$ ratio of H$\alpha$ line profile.}

\section{5. CONCLUSION}

In this article we presented analysis of the H$\alpha$ profiles of the star 60 Cygni. We studied its $EW$(H$\alpha$), $V/R$ and $(V+R)/2$ variation of the profiles.
\par As assumed $EW$(H$\alpha$)  value is related to the size of the disk. Due to this assumption we can say that the formation of the disk takes longer time than the disk extinction (extinction is much steeper than the formation). We gained the period of one ``life'' cycle of the disk to be $\sim6050$ days. Also the last emission activity of the star is stronger than the previous one.
\par We found that there is no evident period of changes in $V/R$ variation, it can be due to the fact that there is no one-armed oscillation of the disk present or it is very small and does not influence the $V/R$ variations.
\bigskip\\
\noindent{\it Acknowledgment} - This work was supported by the M\v{S}MT grant LG15010. This work is based on data from Perek telescope. We would like to thank Ji\v{r}\'{\i} Kub\'{a}t for reading this manuscript which helped us to improve it. We would like to also thank our collegues from Ond\v{r}ejov Observatory who let us use their observed spectra. 

\references

BeSS Database.: 2017, http://basebe.obspm.fr, accessed: 2017-01-04

Carroll, B. W. and Ostlie, D. A.: 2007, in ``Introduction to Modern Astrophysics'', eds. Black, A. R. S., Pearson, San Francisco

Draper, Z.~H. et~al.: 2011, \journal{Astrophys. J.}, \vol{728}, L40

Draper, Z.~H. et~al.: 2014, \journal{Astrophys. J.},  \vol{786}, 120

Hoffleit, D. and Jaschek, C.: 1982, The Bright Star Catalogue,New Haven, Yale University Observatory.

Kogure T. and Leung K.-C., eds, 2007, The Astrophysics of Emission-Line Stars, volume  \vol{342} of Astrophysics and Space Science Library.

Koubsk{\'y}, P. et~al.: 2000, \journal{Astron. Astrophys.},  \vol{356}, 913

Lenz, P. and Breger, M.: 2005, \journal{Communications in Asteroseismology}, \vol{146}, 53

\v{S}koda, P. 1996,: in Astronomical Society of the Pacific Conference Series,  \vol{101}, Astronomical Data Analysis Software and Systems V, ed. G.~H. Jacoby and J.~Barnes, 187

Stellingwerf, R.~F.: 1978, \journal{Astrophys. J.},  \vol{224}, 953

Wisniewski, J.~P. et~al.: 2010, \journal{Astrophys. J.},  \vol{709}, 1306

\endreferences
\end{multicols}
\noindent\parbox{\columnwidth}{
\vskip1cm
{\bf Table~1.} $EW$(H$\alpha$), $V/R$ and $(V+R)/2$ values for spectra observed at the Ond\v{r}ejov observatory.
\vskip.25cm \centerline{
\begin{tabular}{@{\extracolsep{0.0mm}}lllll@{}}
  \hline
 Observers &HJD - 2450000  & $EW$(H$\alpha$) [$\mathrm{\AA{}}$] &$V/R$&$(V+R)/2$\\
\hline
Kor\v{c}\'{a}kov\'{a}, \v{R}ezba&2741.586 &3.633&1.000&1.003\\
Kor\v{c}\'{a}kov\'{a}, Tlamicha&2742.561&3.491&1.004&1.005 \\ 
\v{S}lechta, Kalas& 2835.402&3.464&1.000&1.007 \\
Budovicova, Kotkov\'{a}&2844.423 &3.534&1.003&1.008\\
Wolf, Tlamicha&2900.333 &3.817&0.997&1.005 \\
Koubsk\'{y}, \v{R}ezba& 2955.386& 2.941&0.992&1.005\\
Dovciak, \v{R}ezba& 2983.323& 3.988&1.004&1.003\\ 
Koubsk\'{y}, Tlamicha&3124.434& 2.12&1.012& 1.069\\
Koubsk\'{y}, \v{R}ezba& 3208.403& 3.856&0.998&1.000\\
Koubsk\'{y}, Kalas&3335.310&3.852&0.997&0.998\\
\v{S}lechta, Fuchs&3463.567& 2.015&0.979& 1.026\\
Koubsk\'{y}, Sloup& 3472.478&2.842&1.003& 1.007\\
Kor\v{c}\'{a}kov\'{a}, \v{R}ezba&3511.580&3.057&1.000&1.004 \\
Koubsk\'{y}, \v{R}ezba&3517.398& 3.455&0.991& 1.004\\ 
\v{S}lechta, \v{R}ezba&3613.407& 4.08&1.003&1.005\\
\v{S}lechta, Kotkov\'{a}&3615.498& 4.169&1.003& 1.007\\ 
\v{S}lechta, Sloup&3619.400& 3.296&0.992& 1.009\\
\v{S}lechta, Kotkov\'{a}&3623.444 &2.835&0.997& 1.020\\
Koubsk\'{y}, Sloup&3633.393 &0.848&1.017&1.057  \\
\v{S}lechta, \v{R}ezba&3655.368& 0.494&1.003& 1.121\\
\v{S}lechta, Tlamicha&3658.301& 0.574&1.007&1.101 \\
-&3660.290 &0.709&1.008&1.098 \\
Kor\v{c}\'{a}kov\'{a}, Tlamicha&3672.199& 1.359&0.979& 1.092\\
\v{S}lechta, Votruba&3675.326&0.992&0.993&1.078\\
Kub\'{a}t, Tlamicha&3713.234& 0.651&1.013&1.109\\
Koubsk\'{y}, Sloup&3745.331& 0.854&0.983& 1.057\\
\v{R}ezba&3857.503& 2.499&0.999&1.011\\ 
Votruba, Fuchs&4174.563&-0.348&0.991&1.174 \\
\v{S}lechta, Fuchs&4195.631& -0.159&0.979& 1.172\\
Kor\v{c}\'{a}kov\'{a}, Fuchs&4275.536& 0.485&0.986& 1.160 \\
\v{S}lechta, Fuchs&4594.509& -1.165&1.014&1.246 \\
Netolick\'{y}, \v{R}ezba&4761.302& -0.804&0.992& 1.240 \\
Koubsk\'{y}, Kotkov\'{a}&5040.341&-4.236&0.994& 1.427 \\
\v{S}lechta, Sloup&5279.565& -7.615&1.004&  1.627 \\
\v{S}lechta, Tlamicha&5289.606& -7.531&1.015&1.628 \\
Polster, Kotkov\'{a}&5346.512& -8.008&0.999&1.696\\
Zasche,\v{R}ezba& 5405.430& -7.105&0.983&1.633\\
Koubsk\'{y}, \v{R}ezba&5673.429& -9.432&1.005&1.826\\
 \hline
\end{tabular}}
} \vskip.5cm

\parbox{\textwidth}{
{\bf Table~2.} $EW$(H$\alpha$), $V/R$ and $(V+R)/2$ values for spectra adopted from Bess Database (2017).
\vskip.25cm \centerline{
\begin{tabular}{@{\extracolsep{0.0mm}}lllll@{}}
  \hline
    Observers &HJD - 2400000& $EW$(H$\alpha$) [$\mathrm{\AA{}}$]&$V/R$&$(V+R)/2$\\
\hline
 Desnoux& 49922.451&	-1.199&0.998&1.178\\
Buil& 51327.608	&-8.477&1.003&1.620\\
Buil& 52049.643	&3.276&0.994&1.014\\
Buil& 52127.509	&3.685&0.996&1.008\\
Buil& 52129.488	&3.429&0.986&1.010\\
Buil& 52135.387	&3.863&0.991&1.009\\
Buil& 52212.244	&2.579&0.991&1.012\\
Buil& 52389.622	&2.861&0.996&1.016\\
Buil& 52425.502	&2.890&0.996&1.016\\
Buil& 52760.682	&2.227&0.995&1.022\\
Buil& 53153.618	&2.613&0.997&1.011\\
Thizy, Ribeiro& 54384.446	&-2.064&0.998&1.270\\
Ribeiro& 54405.479	&-1.702&0.985&1.255\\
Ribeiro& 54623.567	&-2.086&0.979&1.300\\
Ribeiro& 54666.499	&-1.736&0.970&1.265\\
Terry& 55066.402	&-7.099&1.050&1.579\\
Desnoux&55066.506	&-5.102&1.010&1.454\\
Guarro Fl\'{o}&55067.406	&-5.479&0.977&1.456\\
Mauclaire&55373.587	&-9.121&0.984&1.808\\
Mauclaire&55381.621	&-8.384&0.977&1.789\\
Mauclaire&55382.566	&-7.659&0.969&1.714\\
Terry&55426.406&	-7.578&0.972&1.656\\
Terry&55745.492&	-7.957&0.996&1.755\\
Garrel&55790.545&	-9.584&1.058&1.896\\
Graham&55846.539&	-8.316&1.016&1.856\\
Desnoux&55858.274&	-8.427&0.988&1.794\\
Graham&55876.546&	-9.404&0.989&1.864\\
Garrel&56112.486&	-10.138&1.003&1.928\\
Garrel&56119.402&	-9.986&1.008&1.943\\
Graham&56187.564&	-10.130&0.962&1.925\\
Ubaud&56500.508	&-11.060&1.013&2.071\\
Desnoux&56506.412&	-12.188&1.003&2.101\\
Graham&56579.619&	-12.640&0.982&2.186\\
Desnoux&56874.393&	-10.814&0.983&2.140\\
Graham&56917.596&	-10.992&0.992&2.128\\
Buil&56934.320&	-10.512&1.011&2.151\\
Pujol&56934.332&	-10.656&1.018&2.138\\
Graham&56938.615&	-9.932&1.010&2.070\\
Terry&57218.425&-7.922&1.024&1.947\\
Graham&57306.519&-8.811&1.015&1.867\\
Houpert&57514.464&-8.2&1.007&1.932\\
Terry&57655.399&-6.437&0.983&1.727\\
\hline
   \end{tabular}
} }

\end{document}